# Pareto Pairwise Ranking for Fairness Enhancement of Recommender Systems

Hao Wang
haow85@live.com
Ratidar.com
Beijing, China

## ABSTRACT

Learning to rank is an effective recommendation approach since its introduction around 2010. Famous algorithms such as Bayesian Personalized Ranking and Collaborative Less is More Filtering have left deep impact in both academia and industry. However, most learning to rank approaches focus on improving technical accuracy metrics such as AUC, MRR and NDCG. Other evaluation metrics of recommender systems like fairness have been largely overlooked until in recent years. In this paper, we propose a new learning to rank algorithm named Pareto Pairwise Ranking. We are inspired by the idea of Bayesian Personalized Ranking and power law distribution. We show that our algorithm is competitive with other algorithms when evaluated on technical accuracy metrics. What is more important, in our experiment section we demonstrate that Pareto Pairwise Ranking is the most fair algorithm in comparison with 9 other contemporary algorithms.

**Keywords:** Pareto Pairwise Ranking, learning to rank, fairness, power law distribution, Bayesian Personalized Ranking

## 1. INTRODUCTION

Recommender system is one of the most successful technologies in our age. By recommending interesting products to users based on predicted user preferences, websites are capable of enlisting a significant increased volume of user traffic to their websites. Without recommender systems, in order to increase user traffic or sales, websites need to spend lavishly on online marketing such as Google Ads. The amount of money is prohibitive to most small-to-medium sized firms. Thanks to recommendation technology, companies could hire a small technical team to solve the user acquisition problem. The cost of such team and its associated hardware facilities is usually a very small fraction of the marketing cost without such team.

Increasing the technical accuracy of recommender system is the major task for most recommender system designers. However, other problems related to recommendation have also raised awareness in the academia and industry. One of such problems is fairness - namely, the algorithmic results are biased towards a specific group of users or products. To understand the severity of this problem, imagine we are building a health recommender system to recommend doctors to patients. If the system is built upon *Collaborative Filtering* , then mild symptoms such as cold and fever would appear everywhere in the similarity computation among users, so that the recommendation results would be contaminated. In addition, common symptoms would have a lot more data than rare diseases. As a consequence, recommendation results for rare diseases would be much worse than mild and common symptoms. Since health industry is often a life or death business, small problems in recommendation results could lead to sever consequences. Therefore fairness in recommender systems is an important research topic, and it's not confined in the health-care industry.

Fairness has largely been researched in the field of learning to rank by global researchers. SIGIR is one of the most popular research venues for the publication of learning to rank based fairness algorithms. Meanwhile, other researchers have used matrix factorization as the research benchmark for fairness research. Matrix factorization-based approaches such as Focused Learning, MatRec, KL-Mat, and Zipf Matrix Factorization were invented recently to tackle the fariness problem.

In this paper, we follow the line of learning to rank based fair algorithms to propose a new algorithm named Pareto Pairwise Ranking that performs the best among 9 modern day recommender systems. We prove that by taking advantage of power law distribution and reformulating *Bayesian Personalized Ranking*, we are able to produce an algorithm that is supreme in fairness metrics and competitive in technical accuracy metrics.

## 2. RELATED WORK

Recommender system has evolved through the following stages: *Collaborative Filtering [1][2] , Matrix Factorization [3], Linear Models [4][5], Hybrid Models [6][7], Learning to Rank [8][9][10], and Deep Neural Networks [11][12][13]*. What is most relevant to our research in this paper is the learning to rank technologies. Learning to Rank technologies (LTR) can be classified into point-wise LTR, pair-wise LTR, and list-wise LTR. Representative approaches include Bayesian Personalized Ranking [8], which optimizes the AUC metric; Collaborative Less is More Filtering [9], which is a list-wise LTR approach. Other algorithms such as GAPfm [10] relies on optimization of different metrics to achieve the optimal ranking list of recommendation results.

One of the intrinsic problems of recommender systems is the fairness problem. The problem can be resolved by adding a regularization term to the loss function of classic models. E. Chi [14] proposed a matrix factorization-based approach with a fairness penalty term in 2017. H. Wang introduced Zipf Matrix Factorization [15] and KL-Mat [16] with different fairness penalty terms. Other than matrix factorization models, learning to rank paradigm is also a very popular choice in fairness modeling [17][18][19].

Power law distribution (Pareto distribution) is ubiquitous in internet data. In recent years, algorithms such as Zipf Matrix Factorization [15], RankMat [20] and Extremal GloVe [21] all rely on properties of power law distribution for algorithmic design. In this paper, the concept of power law distribution is also crucial in our research.

## 3. PARETO PAIRWISE RANKING

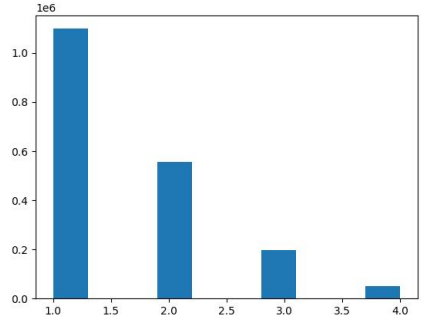

Fig. 1 Positive differences between user item rating value pairs on LDOS-CoMoDa dataset

We observe that the user item rating values of recommender system's input data follow power law distribution. By more experiments (Fig. 1), we also observe that the positive differences between user item rating value pairs also follow power law distribution. To take advantage of the discovery, we define the loss function of our new algorithm (We call it Pareto Pairwise Ranking) as follows :

$$L = \sum_{i=1}^{n} \sum_{j=1}^{m} \sum_{k=1}^{m} P(R_{i,j} > R_{i,k}) I(R_{i,j} > R_{i,k}) \qquad (1)$$

where $R_{i,j}$ is the user item rating value that user i gives to item j, and I is the identity function.

Due to power law distribution, Formula (1) can be redefined in the following way :

$$L = \sum_{i=1}^{n} \sum_{j=1}^{m} \sum_{k=1}^{m} \frac{1}{(U_i^T \cdot V_j - U_i^T \cdot V_k)^\alpha} I(R_{i,j} > R_{i,k}) \qquad (2)$$

Taking the natural log of L, we rewrite the loss function in Formula (2):

$$L = -\alpha \sum_{i=1}^{n} \sum_{j=1}^{m} \sum_{k=1}^{m} \log(U_i^T \cdot V_j - U_i^T \cdot V_k) I(R_{i,j} > R_{i,k}) \qquad (3)$$

We choose to use Stochastic Gradient Descent (SGD) to optimize L for U and V while assuming α is a constant:

$$U_i = U_i + \gamma \alpha \frac{V_j - V_k}{U_i^T \cdot V_j - U_i^T \cdot V_k} \qquad (4)$$

$$V_j = V_j + \gamma \alpha \frac{U_i}{U_i^T \cdot V_j - U_i^T \cdot V_k} \qquad (5)$$

$$V_k = V_k - \gamma \alpha \frac{U_i}{U_i^T \cdot V_j - U_i^T \cdot V_k} \qquad (6)$$

From the formulas above, we notice that the computation of the optimal parameters does not require the knowledge of historic data. However, the parameters are only computed when positivity conditions shown by the identity function are satisfied. Therefore historic data does play a vital role in the computation of parameters.

The complete workflow for Pareto Pairwise Ranking is illustrated as follows:

```
Function Pareto-Pairwise-Ranking :

1    Read input data into the user item rating matrix R
2    For iter in 1: max_iter_number:
3        User_sample = sample users from R
4        α = 1.0
5        For user i in user_sample:
6            U = random sample from uniform distribution
7            Item_sample = sample items from user i's item rating list
8            Item_list = sorted item samples in decreasing rating values
9            V = random sample from uniform distribution
10           For j in 1: max_index_Item_list -1:
11               For k in j+1: max_index_Item_list:
12                   If R[i, j] > R[i, k]:
13                       U_i = U_i + γα (V_j - V_k) / (U_i^T·V_j - U_i^T·V_k)
14                       V_j = V_j + γα U_i / (U_i^T·V_j - U_i^T·V_k)
15                       V_k = V_k - γα U_i / (U_i^T·V_j - U_i^T·V_k)
16   Reconstruct R by dot products of U and V
```

## 4. EXPERIMENT

We test Pareto Pairwise Ranking with 9 other algorithms on MovieLens 1 Million Dataset and LDOS-CoMoDa dataset. The computing platform of our research is a Lenovo laptop with 16 GB and Intel Core i5 CPU. Fig. 2 illustrates the competitiveness of Pareto Pairwise Ranking evaluated using MAE score while Fig. 3 demonstrates the superiority of the algorithm over all of the other approaches.

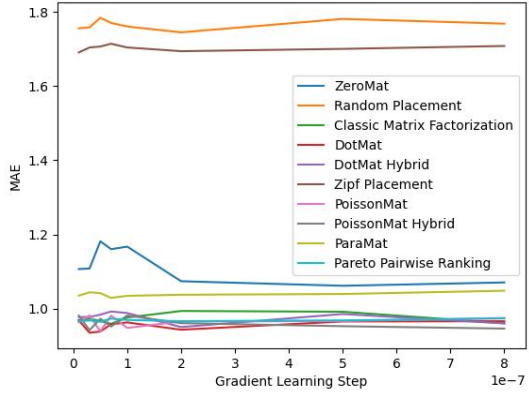

Fig. 2 MAE Comparison

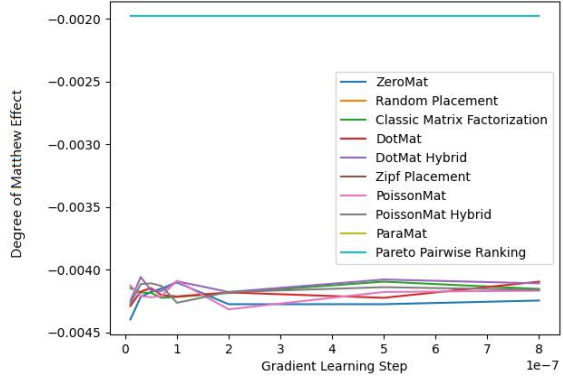

Fig. 3 Fairness Comparison

The 10 algorithms in comparison are ZeroMat [22], Random Placement (recommend randomly), Classic Matrix Factorization, DotMat / DotMat Hybrid [6], Zipf Placement (recommend by power law), PoissonMat / PoissonMat Hybrid [23], ParaMat [24], and Pareto Pairwise Ranking. Pareto Pairwise Ranking ranks No. 1 in fairness comparison [15] evaluated by Degree of Matthew Effect on MovieLens 1 Million Dataset.

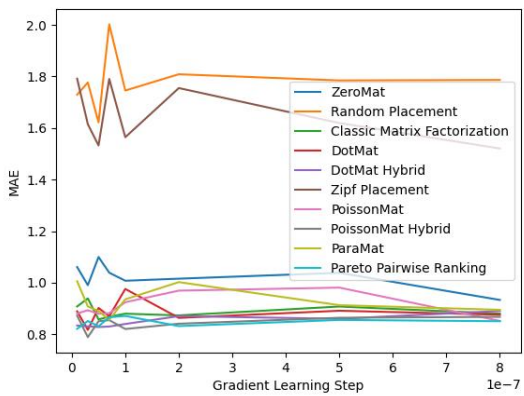

Fig. 4 MAE Comparison

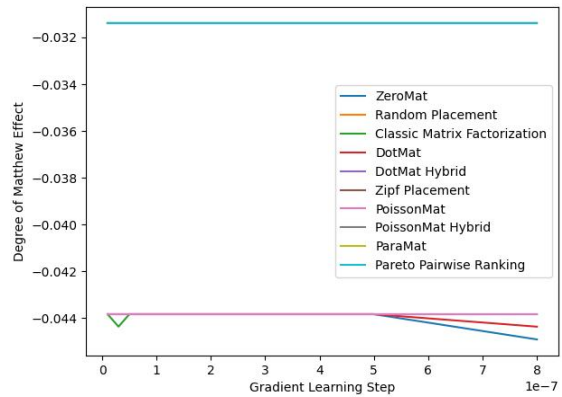

Fig. 5 Fairness Comparison

Fig. 4 and Fig. 5 demonstrates the comparison results on LDOS-CoMoDa dataset. Once again, the result shows that Pareto Pairwise Ranking is competitive on MAE score and supreme on fairness metric [15].

# 5. CONCLUSION

In this paper, we proposed a new learning to rank algorithm named Pareto Pairwise Ranking. We maximize the pairwise relations' maximum likelihood estimator to compute for the optimal parameters, which are dot products of user and item feature vectors. Our innovative idea was inspired by *Bayesian Personalized Ranking* and *Power Law Distribution*. In our experiments, we show that our approach is supreme on fairness metric and competitive on technical accuracy metric.

In future work, we would like to explore the possibility of using *Power Law Distribution* to reformulate other learning to rank algorithms. We wish we could discover more interesting algorithms that can solve the technical accuracy and fairness problems at the same time.